\documentclass[9pt,twocolumn,twoside]{osajnl}

\journal{optica} 
\usepackage{subcaption}

\setboolean{shortarticle}{false}

\title{Intensity interferometry-based 3D imaging}

\author[1,5,*]{Fabian Wagner}
\author[1]{Florian Schiffers}
\author[2]{Florian Willomitzer}
\author[1,2]{Oliver Cossairt}
\author[3,4]{Andreas Velten}

\affil[1]{Department of Computer Science, Northwestern University, 2233 Tech Drive, Evanston, Illinois 60208, USA}
\affil[2]{Department of Electrical and Computer Engineering, Northwestern University, 2145 Sheridan Road, Evanston, Illinois 60208, USA}
\affil[3]{Department of Biostatistics and Medical Informatics, University of Wisconsin-Madison, 610 Walnut Street, Madison, Wisconsin 53726, USA}
\affil[4]{Department of Electrical and Computer Engineering, University of Wisconsin-Madison, 1415 Engineering Drive, Madison, Wisconsin 53706, USA}
\affil[5]{Department of Computer Science, Friedrich-Alexander-Universität, Martensstrasse 3, Erlangen, Bavaria 91058, Germany}

\affil[*]{Corresponding author: fabian.wagner@fau.de}

\begin{abstract}
The development of single-photon counting detectors and arrays has made tremendous steps in recent years, not the least because of various new applications in, e.g., LIDAR devices. In this work, a 3D imaging device based on real thermal light intensity interferometry is presented. By using gated SPAD technology, a basic 3D scene is imaged in reasonable measurement time. Compared to conventional approaches, the proposed synchronized photon counting allows using more light modes to enhance 3D ranging performance. Advantages like robustness to atmospheric scattering or autonomy by exploiting external light sources can make this ranging approach interesting for future applications.
\end{abstract}

\begin{document}

\maketitle

\section{Introduction}

\begin{figure}[h!]
\centering\fbox{\includegraphics[width=\linewidth]{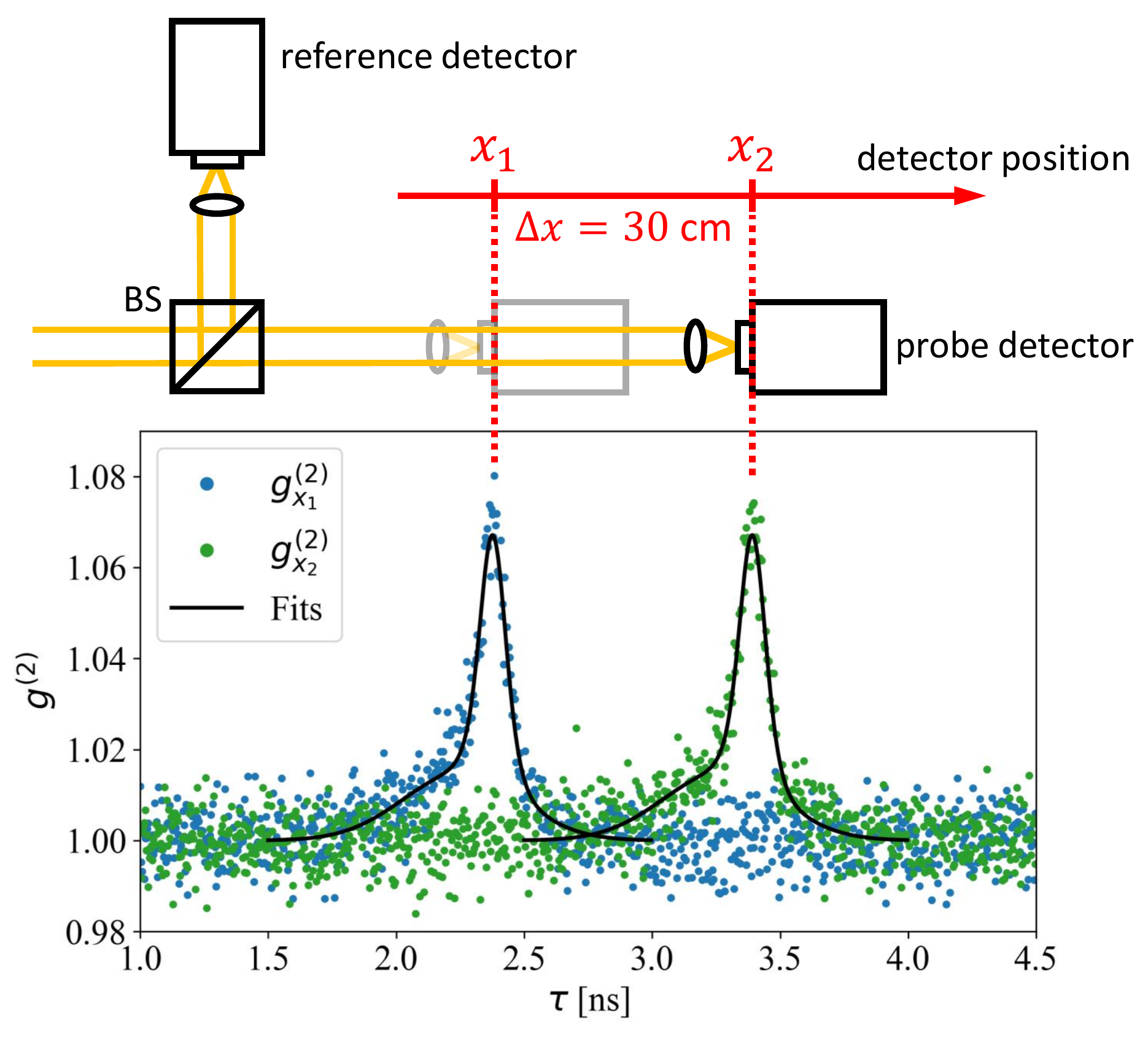}}
\caption{Sketch of the intensity interferometry-based ranging principle. The position of the measured intensity correlation peak shifts proportional to the length of the interferometer probe arm ($\Delta\tau \sim \Delta x = x_2-x_1 = 30\,\text{cm}$), assuming a fixed reference detector position. The constant of proportionality is given by the speed of light $c$.}
\label{fig:rangingPrinciple}
\end{figure}

3D imaging has uses for industrial applications in different fields, e.g., aerospace, automotive, and medical imaging, or for autonomous navigation \cite{foix2011lock, morimoto2020megapixel, Willo3DCam17, Harendt14, antipa2018diffusercam, Huber11, lidarSelfDriving, GerdWhy, Caulier:10}. In this paper, we present the first intensity interferometry-based 3D imaging system, exploiting the inherent photon bunching signature of thermal light. We demonstrate that we can 3D image a basic scene without any hardware control over the light source. Furthermore, we introduce a novel photon counting technique using gated \textit{single-photon avalanche diode} (SPAD) technology. Synchronizing two SPAD detectors allows us to leverage the system's acquisition performance and perform a proof-of-principle 3D scan.\\
Conventional 3D imaging approaches distinguish between triangulation and time-of-flight-based techniques. Whereas the triangulation angle limits depth resolution, time-of-flight imaging exploits parallel probe and reflection beam paths, and the depth accuracy is determined by the modulation frequency and the detector bandwidth \cite{foix2011lock}. As a result, more flexible scenes, including featureless objects and vast standoff distances, can be 3D imaged, e.g., with LIDAR devices or ToF-cameras. However, $\text{MHz}$ modulation of the utilized laser restricts the depth resolution of intensity-modulated ToF devices to centimeters \cite{lange2001solid, stoppa2010range}. Higher modulation frequencies typically require more sophisticated hardware \cite{yasutomi2013time, li2019mega}. Interferometric approaches allow distance measurements down to nanometer resolution but are limited in their measurement depth range due to the periodic nature of light waves \cite{steel1983interferometry}. In addition, small distortions on the order of the wavelength lead to decorrelation and prohibit ranging in uncontrolled environments like turbulent atmosphere or through scattering media. Recent works present techniques using picosecond modulation frequencies combined with single-photon detection \cite{liu2019non} or optical multi-wavelength heterodyning approaches \cite{willomitzer2019high, willomitzer2019synthetic} to exploit modulation frequencies larger than used in conventional intensity-modulated ToF but smaller than optical frequencies. This allows imaging through scattering media and looking around corners with sub-mm depth precision.\\
In summary, ToF techniques require active modulation of the utilized light source in order to temporally encode the probe beam. This poses the question if an intensity-modulated ToF device could also operate, only using ambient light, without hardware control over the light source. It turns out that the inherent intensity fluctuations of thermal light on pico- to femtosecond timescales result in a high-frequency intensity modulation, which can be used for sub-millimeter ranging, as we demonstrate in this paper. Recording light intensity correlations with a fixed reference detector enables depth measurements, as illustrated in Fig. \ref{fig:rangingPrinciple}. Imaging techniques exploiting incoherent light are an attractive option compared to laser illumination because no artificial light source is required, the 3D sensing system consumes less power, and is nearly impossible to detect. So far, intensity interferometry with thermal light has been almost exclusively used for stellar imaging purposes. However, the development of single-photon counting detectors and arrays has made tremendous steps in recent years \cite{morimoto2020megapixel, ulku2018512, ceccarelli2018fully}, such that further applications, like intensity interferometry-based 3D imaging, are now conceivable.

\subsection{Main contributions}
The following bullet points highlight our main contributions:
\begin{itemize}
  \item To the best of our knowledge, we are the first to build a 3D imaging system based on intensity interferometry only using a thermal light source.
  \item We present a first example scan, 3D imaging a basic scene with a depth resolution of $\Delta d = 1.7 \,\text{mm}$. In principle, our system can achieve sub-millimeter depth precision ($\Delta d = 0.3 \,\text{mm}$ is experimentally demonstrated), independently from the standoff distance.
  \item We propose a novel photon counting technique, synchronizing the two SPAD detectors by using gated SPAD technology in order to leverage the performance of our 3D imaging intensity interferometer. The approach improves our acquisition time by a factor > 2.
\end{itemize}

\subsection{Spatial intensity correlations}
In 1956 Hanbury Brown and Twiss measured the angular diameter of the star Sirius with a $10\,\text{m}$ baseline intensity interferometer exploiting the degree of second-order coherence of light \cite{brown1956test}. In a previously conducted tabletop experiment, they demonstrated that not only electric fields of a light source are correlated during their coherence time but also their intensity fluctuations \cite{brown1956correlation}. Roy Glauber provided the mathematical description of the underlying coherence theory seven years later \cite{glauber1963quantum}. It includes the quantum description of optical coherence, for which he won the Nobel Prize for in 2005. His theories set the foundation for quantum optics and are still in use today.\\
In the 1960s and 1970s, Hanbury Brown determined the angular diameter of 32 stars with his Narrabri stellar intensity interferometer in Australia \cite{hanbury1974angular}. One advantage of (second-order) intensity interferometry over (first-order) Michelson stellar interferometry is the robustness against atmospheric turbulences \cite{tan2016optical}. The propagation length difference of light detected in the two interferometer arms disturbs the correlated signals. Whereas in a Michelson interferometer a perturbation of the phase of the electric field on the order of the wavelength is sufficient to suppress correlations, intensity interferometry is robust to random signal changes on the order of the coherence length, which can be considerably larger. However, the invention of adaptive optics in the 1990´s replaced the need for  intensity interferometers in astronomy, due to the superior light sensitivity of Michelson interferometers \cite{davis1999sydney}.\\
With recent developments towards broad bandwidth, high quantum efficiency photon detection, and high-speed signal processing electronics, the construction of large-baseline intensity interferometer becomes tempting for stellar imaging purposes \cite{pilyavsky2017single, horch2013intensity, dravins2015long}. As intensities instead of electric fields interfere, the detection of light with multiple telescopes with subsequent offline signal correlation is much easier to realize and allows a simpler implementation of very large baseline interferometers (e.g., the Cherenkov Telescope Array) \cite{kieda2019astro2020, kieda2019augmentation, le2006optical}. Experimental small scale proof of principle was already demonstrated \cite{weiss2018stellar, guerin2017temporal, zmija2020led}.\\
In addition to astronomical applications, intensity interferometry has also been demonstrated in terrestrial approaches. Measuring the spatial coherence function of entangled photons enables 2D imaging of binary structures, called lensless ghost imaging. A shadow image can be acquired with the pixel "brightness" given by the degree of second-order coherence \cite{liu2014lensless, defienne2020quantum, boiko2009quantum}.

\subsection{Temporal intensity correlations}
Since the depth resolution of conventional time-of-flight cameras is restricted to the modulation frequency of the light source, their depth accuracy is usually on the order of centimeters \cite{mccarthy2009long}. In contrast, the inherent modulation of thermal light provides a high-frequency signal that, in principle, can be used to perform ranging with sub-millimeter accuracy. In this paper we introduce an intensity interferometry-based device that only uses the inherent fluctuations of real thermal light to perform time-of-flight imaging and measure the topography of a real 3D object. \textit{To the best of our knowledge, we are the first to build a 3D imaging system based on intensity interferometry.} The work that comes closest to ours measured the length of glass fibers with micrometer resolution.
Pseudo-thermal light obtained by creating moving spatial speckle using a CW-laser and a rotating ground glass disc has been used to demonstrate ranging \cite{zhu2012thermal}. Pseudothermal light sources are a way to create light with intensity fluctuations that are statistically similar to \cite{mehringer2017optical, schneider2018simulating}, but in the referred ranging approach much slower and more intense than the fluctuations that would be created from a real thermal light source.\\
Other approaches use entangled photons from parametric down-conversion produced by pumping a nonlinear material with a CW-laser. They aim to synchronize two clocks \cite{valencia2004distant} or again to determine the length of a fiber delay \cite{jun2013distance}. Altogether, completely free space distance measurements using real thermal light have not been demonstrated yet.

\section{Theory}
Analog to the visibility of interference fringes of electric field amplitudes $E(t)$, the degree of second-order coherence $g^{(2)}(\tau)$ quantifies the correlation of light intensities $I(t)$. It is defined as the auto-correlation function
\begin{equation}
\label{eq:g2_def}
g^{(2)}(\tau)=\frac{\langle E^*(t)E^*(t+\tau)E(t+\tau)E(t) \rangle}{\langle E^*(t)E(t) \rangle \langle E^*(t+\tau)E(t+\tau) \rangle} = \frac{\langle I(t) I(t+\tau)\rangle}{\langle I(t) \rangle \langle I(t+\tau)\rangle}
\end{equation}
with the relative time delay $\tau$ and brackets $\langle \dots \rangle$ indicating time averaging. For Poissonian light, which can be well represented by the output of a continuous-wave laser with a long coherence time, $g^{(2)}(\tau)$ yields the constant value $1$.\\
In contrast, thermal light sources (e.g., the sun or light bulbs) emit super-Poissonian light with additional intensity fluctuations caused by, e.g., collisions of light-emitting atoms within the source, or due to the Doppler shift. Therefore, they possess a much shorter coherence time $\tau_c$, usually on the order of $\text{fs}$ to $\text{ps}$. When determining $g^{(2)}(\tau)$ of a single spatial mode of such light, one can show that $g^{(2)}=2$ for $\tau=0$. However, for $|\tau| \gg \tau_c$, due to vanishing coherence, $g^{(2)}$ reduces to $1$, resulting in a global correlation peak around $\tau=0$, the so-called \textit{bunching signature} of thermal light \cite{fox2006quantum, loudon2000quantum, mandel1995optical}. The intensity fluctuations, which yield this inherent correlation of light, can be regarded as a high-frequency random intensity modulation.\\
Experimentally, the degree of second-order coherence can be accessed by a Hanbury Brown-Twiss interferometer, which constitutes the basis for most intensity interferometers. Using single-photon counting detectors, the numerator of Eq. \ref{eq:g2_def} $\langle I_1(t) I_2(t+\tau)\rangle =:G^{(2)}(\tau)$ can be measured by recording photon arrival times of a light mode, split and detected in two interferometer arms 1 and 2. Counting photon arrival time differences $\tau$ in a histogram yields a correlation peak of width $\tau_c$ and position $\tau_0$, due to the increased probability of detecting two photons---one in each arm---within the coherence time of thermal light. $\tau_0$ is dependent on the pathlength difference of the interferometer arms. $g^{(2)}(\tau)$ can be calculated by normalizing $G^{(2)}(\tau)$ by its constant baseline for $|\tau-\tau_0| \gg \tau_c$.\\
Due to high quantum efficiency (up to $50\,\%$) and low timing jitter (tens of $\text{ps}$), it is expedient to use \textit{single-photon avalanche diodes} (SPADs) in intensity interferometry experiments. However, SPADs possess a dead time $t_{\text{dead}} \approx 77\,\text{ns}$ after each photon detection. During this dead time they are unable to detect photons, which limits their maximum count rate to $\frac{1\,\text{s}}{t_{\text{dead}}}\,\text{cps}$ ($\text{cps} = \text{counts per second}$), usually on the order of several millions $\text{cps}$. In a conventional intensity interferometer, the SPADs in the two interferometer arms operate independently from each other. Especially when operating the SPADs close to saturation, a significant number of photons is then missed due to the dead time, including correlated photon pairs which would contribute to the photon bunching signature. In order to leverage the performance of our 3D imaging intensity interferometer, we propose a novel photon counting technique, synchronizing the two SPAD detectors by using gated SPAD technology.

\section{Intensity interferometry-based 3D imaging device}
\begin{figure}[h!]
\centering\fbox{\includegraphics[width=\linewidth]{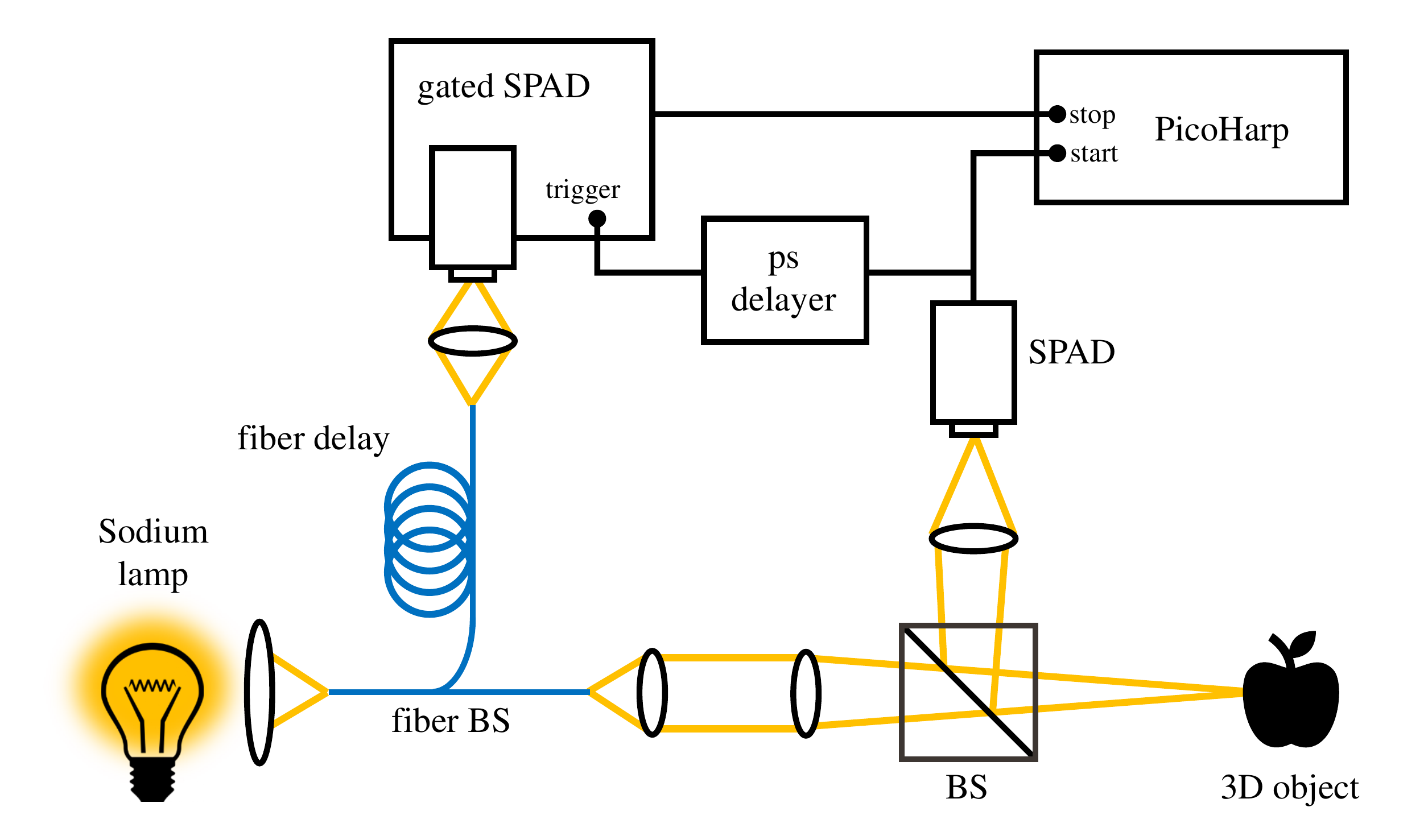}}
\caption{Sketch of the gated confocal 3D imaging setup. A gated SPAD detector is used in the reference interferometer arm. It is only sensitive for photons after being triggered by the conventional SPAD. This detector synchronization makes the setup particularly sensitive for histogram events around the correlation peak.}
\label{fig:gatedSetup}
\end{figure}
\begin{figure}[h!]
\centering\fbox{\includegraphics[width=\linewidth]{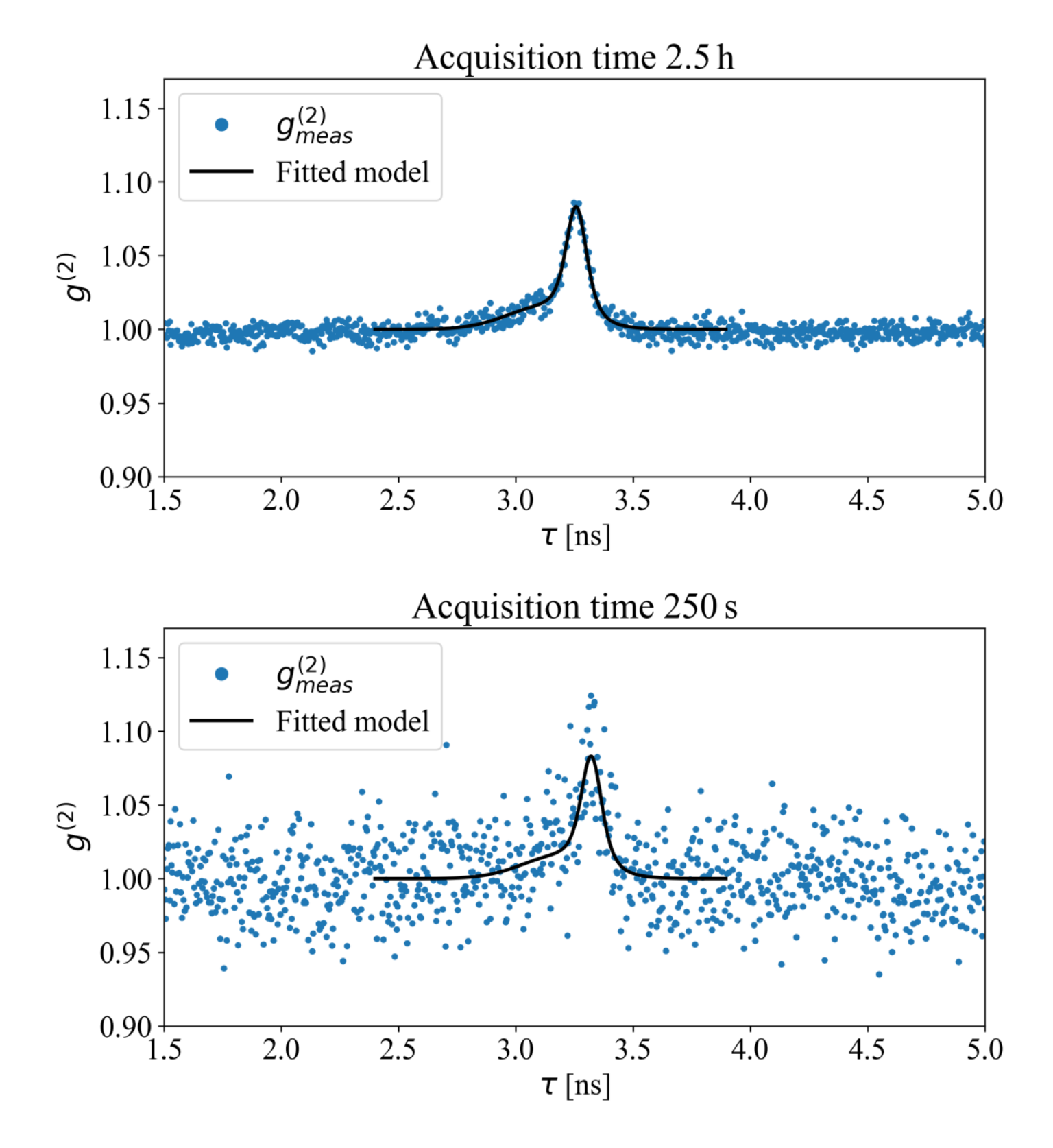}}
\caption{Top: The normalized and calibrated second-order correlation peak from the $2.5\,\text{h}$ calibration measurement is least-square fitted by the developed model function to determine its parameters. A coherence time of $\tau_c = (66 \pm 9)\,\text{ps}$ is determined. Bottom: The correlation histogram from a single raster-scan position, acquired in $250\,\text{s}$.}
\label{fig:g2_longterm}
\end{figure}

We implemented a novel 3D imaging technique that exploits the bunching signature of real thermal light.
The depth-ranging principle of our system works as follows:
To be correlated, photons must originate from the same photon bunch.
Therefore, the position of the correlation peak is dependent on the pathlength difference between the two detector arms that are input to a Time-Correlated Single Photon Counting Device (TCSPC).
We fix the reference arm of the interferometer to provide a static pathlength reference to measure against.
When the pathlength in the sample arm of the interferometer changes, the position of the measured correlation peak shifts proportionally to the change in pathlength divided by the speed of light.
This shift in the correlation peak, illustrated in Fig. \ref{fig:rangingPrinciple}, provides a mechanism to quantify the depth of a scene relative to a fixed reference, enabling 3D range sensing.\\
Our real thermal light source consists of quasi-monochromatic light (mainly the $589.0\,\text{nm}$ and $589.6\,\text{nm}$ emission lines) emitted from a Sodium lamp.
The light is coupled into a single-mode fiber beam splitter to achieve sufficient spatial and temporal coherence.
The outputs of the fiber beam splitter distribute the photons equally into the two interferometer arms.
A fiber coupler is used in the upper---so-called \textit{reference}---interferometer arm to focus the beam onto the sensitive area of a Micro Photon Devices (MPD) FastGatedSPAD (gated SPAD), as illustrated in Fig. \ref{fig:gatedSetup}.
In external trigger mode, the gated SPAD is only sensitive for impinging photons after being triggered by an external signal.\\
The second interferometer arm, also called the \textit{sample arm}, contains the investigated object.
After collimating the fiber output, a 50:50 beam splitter is used to illuminate the object in a confocal arrangement.
Light being reflected by the object reenters the beam splitter and is subsequently collected by an objective lens.
The lens is used to collect the maximum number of reflected photons and focus them onto the sensitive area of an MPD PDM Series SPAD.
In order to measure photon correlations, photon arrival times at the two SPAD detectors are correlated with a Time Correlated Single Photon Counter (TCSPC, PicoQuant PicoHarp300).
The TCSPC continuously determines arrival time differences $\tau$ of the signals provided by the SPAD detectors.
It creates a histogram of $\tau$ events, which---in case of photon bunching---exhibits a correlation peak of the width of the coherence time.\\
Unlike conventional approaches, we present an optimized photon counting in order to maximize the average counts per bin in the histogram region around the correlation peak to leverage the signal-to-noise ratio (SNR) of $g^{(2)}_{\text{meas}}(\tau)$.
The SNR is defined by the correlation peak height divided by the standard deviation of the histogram counts far away from the correlation peak.
Caused by additional optics, as well as the partly diffusive beam reflection on the investigated object, the count rate in the sample arm of the interferometer is significantly lower than in the reference arm, in our case, a factor of $\approx 43$ for a relatively specular reflecting object.
In the experiment, count rates of $\gamma_s = 1\times 10^5\,\text{cps}$ in the sample arm yield $\gamma_r = 4.3\times 10^6\,\text{cps}$ in the reference arm.
Due to the $t_{\text{dead}}=77\,\text{ns}$ dead time of the SPAD detectors after each photon detection, the reference arm SPAD operates close to saturation \cite{MPDSPAD}.\\
Let us assume two correlated photons, originating from a single photon bunch, getting split in the two interferometer arms.
In principle, the photons would be counted as a histogram event, contributing to the correlation peak.
However, caused by the large photon number in the reference arm, there is a considerable probability $p$ that during the arrival of the correlated photon the reference SPAD is in its dead time due to a previously detected photon:
\begin{align}
\label{eq:FastGatedSPAD}
p(\text{"reference SPAD is blind"}) = \frac{\gamma_r \times t_{\text{dead}}}{1\,\text{s} \times  1\,\text{cps}} = 33.11\,\%.
\end{align}
For the PDM SPAD in the sample arm the probability is given by
\begin{align}
\label{eq:PDMSPAD}
\begin{split}
p(\text{"probe SPAD is blind"}) &= \frac{\gamma_s \times t_{\text{dead}}}{1\,\text{s} \times  1\,\text{cps}} = 0.77\,\% \\
&\ll p(\text{"reference SPAD is blind"}).
\end{split}
\end{align}
The proposed detection scheme, containing a gated SPAD, partly retrieves missed interesting $\tau$ events due to saturation of the gated SPAD.
Here, the output of the object arm SPAD is not only used for correlating photon arrival times but also for triggering the gated SPAD, such that it is only sensitive for photons after a photon was detected at the object arm SPAD.
Due to the strongly unbalanced photon count rates in the two interferometer arms, the lower PDM SPAD count rate limits the number of recorded $\tau$ events.
Gating the reference arm SPAD allows to detect preferentially correlated photons in order to maximize the $\tau$ counts---and such the SNR---around the correlation peak in the $g^{(2)}_{\text{meas}}(\tau)$ histogram.\\
In the experiment, an MPD picosecond delayer is used to adjust the SPAD trigger in order to shift the gate window around the correlation peak.
To compensates for the time it takes to open the gate of the reference arm SPAD ($\approx 100\,\text{ns}$), a $25\,\text{m}$ single-mode fiber delay in the reference arm is used.\\
In order to accurately determine the position of the correlation peak, a model of the correlation peak is used.
In reality, $g^{(2)}_{\text{meas}}(\tau)$ consists of $g^{(2)}(\tau)$ convolved with the detector response of both detectors.
For the procedure used to fit the $g^{(2)}_{\text{meas}}(\tau)$ function, the interested reader is referred to \cite{wagner2020intensity, schneider2018correlation}.
Because raster-scanning the object does not change the coherence properties of the light, and therefore the shape of the coherence function remains the same, the model function parameters can be determined by least-square fitting a single high-SNR correlation data set (python scipy).
Figure \ref{fig:g2_longterm} exhibits this fit on data, captured in a $2.5\,\text{h}$ calibration measurement.
Note that the correlation histogram~$g^{(2)}_{\text{meas}}(\tau)$ is normalized by the baseline counts and calibrated with a measurement from a light source holding no photon bunching to remove oscillating electronic artifacts from the TCSPC histogram.

\begin{figure}[h!]
\centering\fbox{\includegraphics[width=\linewidth]{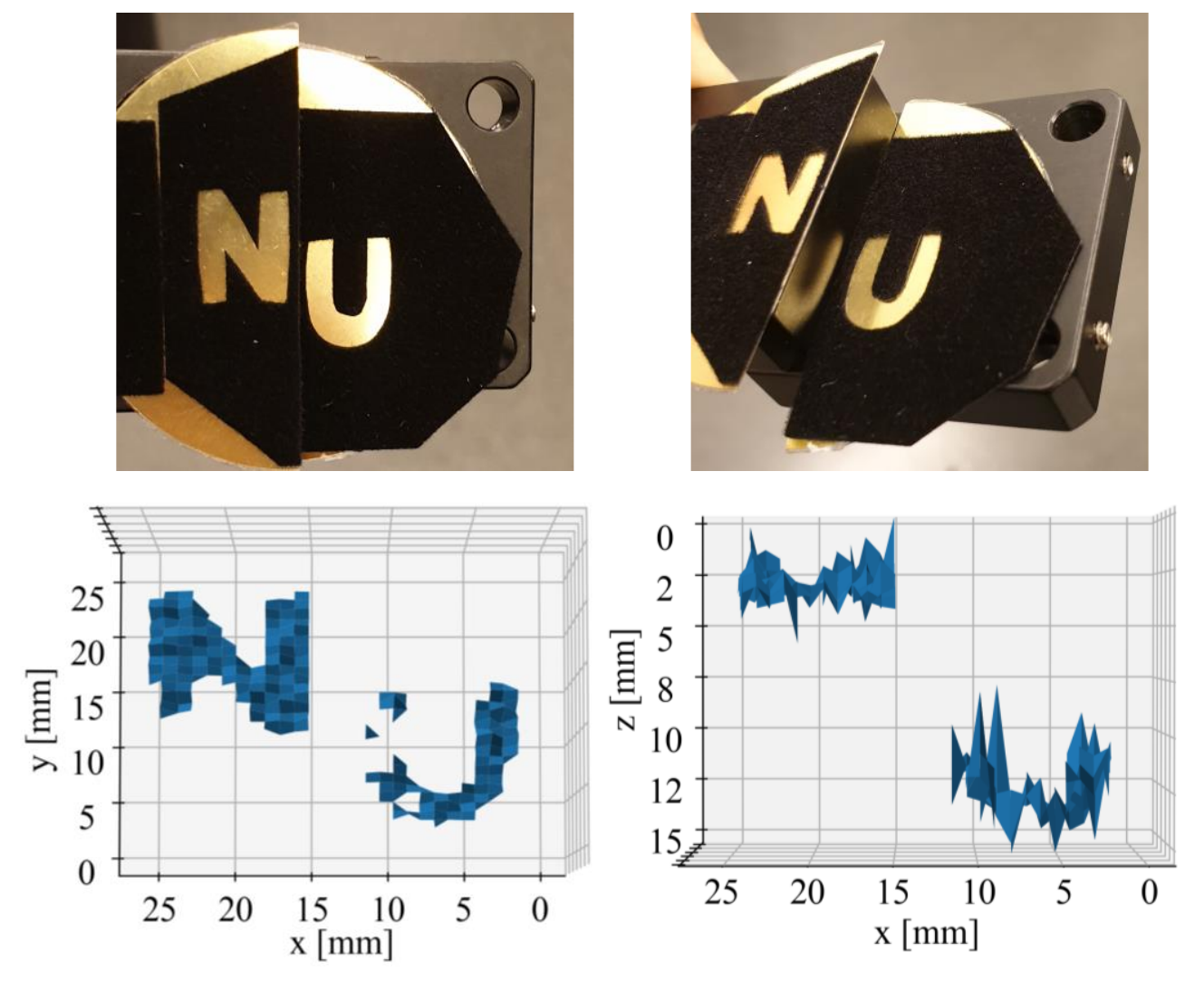}}
\caption{3D scan of two metal-coated letters arranged in different depths $9\,\text{mm}$ apart (see upper photographs). The bottom left-hand plot shows the top view on the scanning region, while the depth information of the scene can be obtained from the bottom right-hand reconstruction, i.e., the side view of the same 3D plot.}
\label{fig:3DScansLetters}
\end{figure}

With the presented setup, a basic 3D object can be raster-scanned by mounting it on a 2D translation stage.
At every pixel position $(x,y)$ the depth information is acquired by recording the correlation function $g^{(2)}_{(x,y), \text{meas}}(\tau)$ and determining the position of the respective correlation peak.
The position of the correlation peak is again determined via least-square fit.
However, this time all parameters of the model function are known by the calibration measurement, except for the peak position $\tau_0$.\\
With this approach, we were able to 3D image a basic scene, consisting of two metal letters arranged in two different depths $9\,\text{mm}$ apart.
The metal surface is highly specular but sufficiently rough to emit a fully developed speckle pattern, reflecting enough photons to keep the measurement time reasonable while still showing proof that a rough surface, which scrambles the wavefront, does not destroy the measured photon bunching signature.
For the presented $30 \times 30$ raster-scan, an acquisition time of $250\,\text{s}$ per pixel is chosen.
The reconstructed volume is shown in Fig. \ref{fig:3DScansLetters}.
Scan positions where no correlation peak could be obtained are excluded.\\
The depth resolution of the presented setup is determined by the uncertainty of the peak position fit parameter, which in turn is determined by the number of detected photons.
In the correlation histogram acquired in a $250\,\text{s}$ scan, displayed in Fig. \ref{fig:g2_longterm}, the uncertainty of the peak position yields $\Delta \tau = 5.8\,\text{ps}$, which corresponds to a depth precision of $\Delta d = 1.7\,\text{mm}$.
However, longer acquisition times improve the histogram statistics and, therefore, can achieve sub-millimeter depth resolution, e.g., the calibration data in Fig. \ref{fig:g2_longterm} with $\Delta \tau = 1\,\text{ps}$ and $\Delta d = 0.3\,\text{mm}$.

\section{Speed improvement of the gated acquisition}
\begin{figure}[h!]
\centering\fbox{\includegraphics[width=\linewidth]{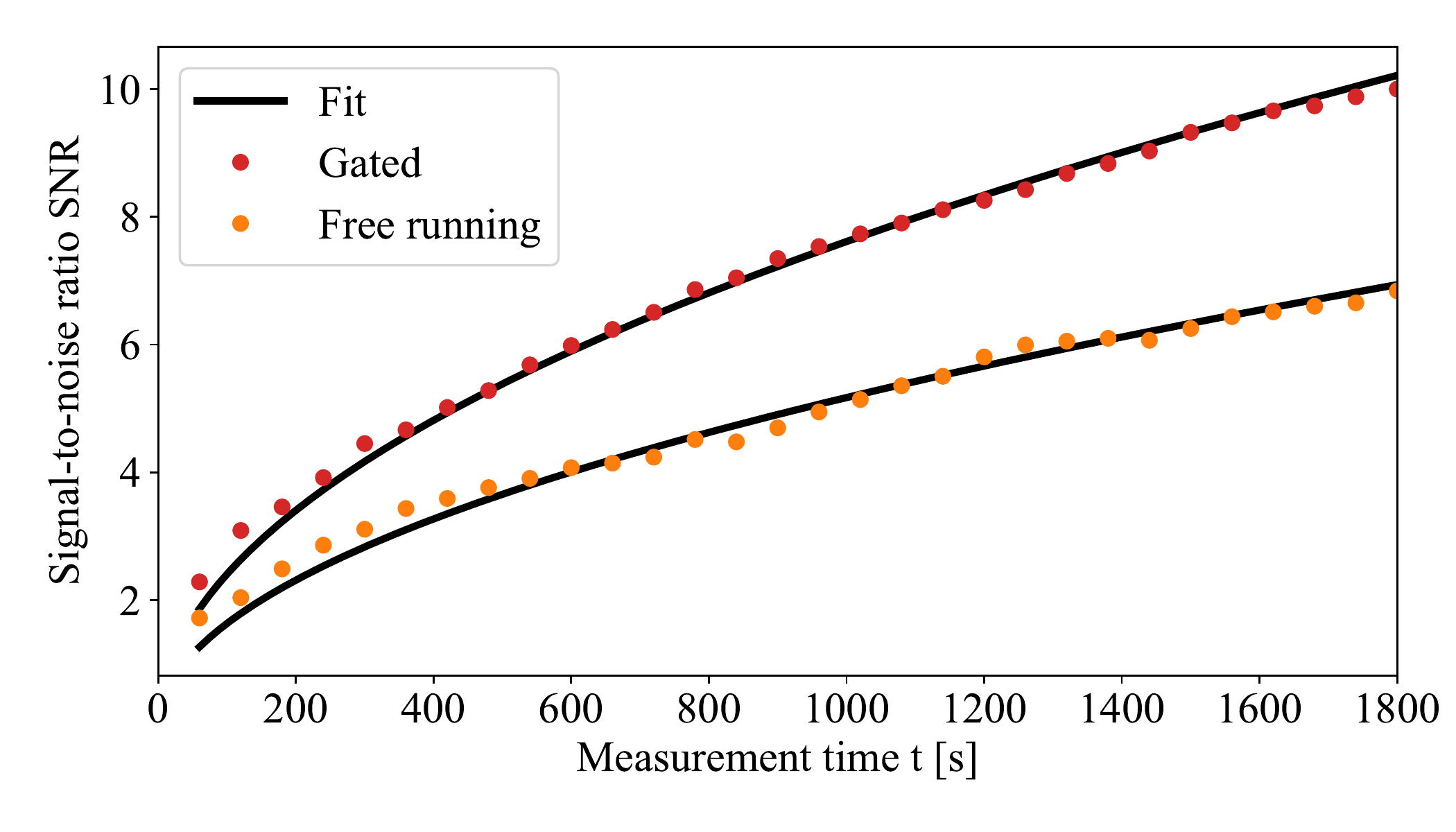}}
\caption{The gated photon counting technique provides an improved correlation peak SNR compared to the conventional approach, where both SPADs record photons independently from each other.}
\label{fig:compare_speed_SNR}
\end{figure}

In order to quantify the speed improvement of the gated photon counting technique compared to the conventional approach, the SNR serves as a quality measure of the correlation function.
The signal of a correlation histogram is determined by fitting the model function to the data and estimating the visibility of the distribution, defined as its height
\begin{equation}
\nu = g^{(2)}(\tau_0)-1.
\end{equation}
Various parameters like histogram bin size or spatial coherence of the light source can decrease the visibility from its maximum value of $2$.
However, for a fixed optical setup, the visibility can be assumed constant, as the detection method and coherence of the light do not change. The respective noise is derived by calculating the standard deviation $\sigma_{g^{(2)}}$ of the baseline fluctuations away (i.e., $|\tau - \tau_0| \gg \tau_c$) from the correlation peak.\\
As SPAD detectors are sensitive to single photons, to every detected photon a particular arrival time is assigned. This photon quantization necessarily leads to fluctuations of the intensity over time and is called shot noise \cite{schottky1918spontane}. The photon arrival time difference $\tau$ directly follows from the measured photon arrival times and therefore underlies the same statistics. Shot noise arises Poissonian distributed such that its standard deviation is given by the square root of the average counts per histogram bin $N$. After normalizing the correlation histogram, it becomes evident that for a constant photon count rate the SNR can be modeled by a function of the type
\begin{equation}
\sigma_{g^{(2)}} \sim \frac{1}{\sqrt{t}} \Rightarrow \text{SNR}_{g^{(2)}} = A \sqrt{t}
\end{equation}
with parameter $A$ and acquisition time $t$. Figure \ref{fig:compare_speed_SNR} demonstrates the correlation peak SNR dependence on the acquisition time, for both the conventional and the gated photon counting technique. Note that the light intensities at both detectors equal (count rates in free running mode $\approx 4.7\times 10^6 \,\text{cps}$ and $\approx 2.2\times 10^5 \,\text{cps}$) for both experiments and only the photon counting method is changed. One recognizes a speed improvement of a factor $>2$ in order to reach, e.g., $\text{SNR}=6$. Note that this particular speed improvement is only valid for the present light levels. However, for more unbalanced photon numbers in the two interferometer arms, the gated counting can produce far more significant SNR gains as it utilizes the sparse probe arm photons with largest efficiency relative to the correlation peak SNR.\\

The gated acquisition technique can handle light intensities, that would saturate the reference arm SPAD in a conventional intensity interferometric approach. Due to the gating, the photon count rate of the reference SPAD can be reduced from $\approx 4.7\times 10^6 \,\text{cps}$ to $\approx 3.5\times 10^4 \,\text{cps}$. Hence, more spatial light modes---which would otherwise saturate the reference arm SPAD---can be used in the experiment, in order to acquire improved photon statistics and such to leverage the acquisition performance.\\
We used our experimental prototype in Fig. \ref{fig:gatedSetup} to simulate acquisition with a multi-mode fiber. To accomplish this, multiple data sets of photon arrival times are captured in subsequent single-mode fiber measurements. Photons from two different modes are always uncorrelated. By computationally cross-correlating an arbitrary number of such independent data sets, light, originating from multiple source points, is imitated.\\
When correlating photon arrival times of multiple optical modes (i.e., acquired data sets), two corrections are applied:
\begin{enumerate}
\item When correlating independent data sets, photon arrival times within the dead times of the SPADs can exist. Such photon events are removed from the data.
\item Due to the gated acquisition technique, the sensitive time windows of the SPADs are synchronized. When now correlating the merged list of photon arrival times, i.e., calculating $\tau = t_2 - t_1$, events where both photon arrival times $t_1$ and $t_2$ originate from the same measurement are more likely to happen. As a result, $\tau$ cross terms with $t_1$ and $t_2$ from uncorrelated modes are underrepresented. To compensate for that, for every measured $t_1$ a random event $t_2^*$ is introduced, which ends the $\tau$ measurement prematurely if $t_2^* < t_2$. The probability of $t_2^*$ is uniformly distributed over all existent histogram time bins between opening the SPAD gate and the respective $\tau = t_2 - t_1$ event and calculated by taking into account the photon count rates of the correlated modes and the SPAD dead time.
\end{enumerate}

\begin{figure}[h!]
\centering\fbox{\includegraphics[width=\linewidth]{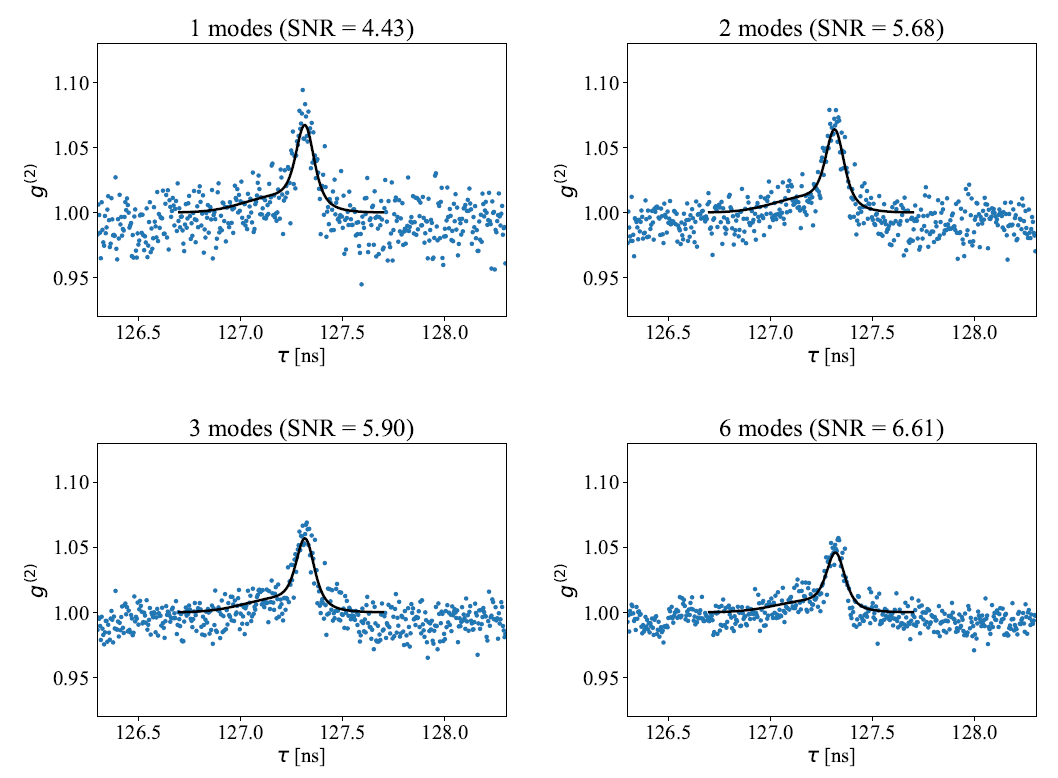}}
\caption{Results of computationally combining gated correlation measurements to imitate multiple light modes coupled into the experiment. The respective title exhibits the number of combined measurements together with the achieved SNR within the constant total acquisition time $t=300\,\text{s}$.}
\label{fig:multiple_modes}
\end{figure}

Figure \ref{fig:multiple_modes} exhibits the correlation results of up to six combined measurements with constant total acquisition time $t=300\,\text{s}$. The results of our multi-mode simulation experiment indicate that, for our setup, the SNR of the correlation peak can be further increased from $4.43$ to $6.61$ by using light from an optical fiber with a core intersection area six times the one of a single-mode optical fiber. Note that the individual correlated single-mode measurements, in reality, consist of more than a perfect single optical light mode as, e.g., different polarization modes are not filtered.\\
When further increasing the number of included light modes, one recognizes a declining SNR again, as then saturation effects of the probe arm SPAD can not be neglected anymore. In general, one can conclude that the SPAD dead time limits the maximum number of included light modes and such the acquisition performance.

\section{Future outlook}
In this work, we demonstrate the first intensity interferometry-based ranging experiment using thermal light and provide novel methods and insight into thermal light measurements using gated SPAD photon counting detectors. While our method improves the required acquisition times, these are still too long to make our approach competitive. A major detrimental factor is the dead time inherent to SPAD detectors and the associated non-linear behavior when correlating two signals. While it should be straight forward to improve the SNR of the measurement by adding more light modes, this would result in photon rates too high for the SPAD to handle.\\
Enhanced hardware with shorter dead times like \cite{ceccarelli2018fully} could reduce the measurement time by faster recording better photon statistics. In addition, dramatically faster acquisition times can be achieved by eliminating scanning and using a SPAD array \cite{morimoto2020megapixel}. Photon arrival times could be counted simultaneously with multiple SPAD pixels in each interferometer arm while exploiting light from multiple light modes.\\
Neglecting the advantages of SPAD detectors like high quantum efficiency and temporal resolution or comparably easy access to the correlation function via TCSPC principle, the dead time limitation could be tackled by using analog detectors in combination with an electronic mixer, as Hanbury Brown and Twiss did when investigating stellar light sources \cite{brown1956test}. Compared to the single-photon counting approach, analog mixing is particularly sensitive for the intensity fluctuations while large background intensities cancel out. This, in principle, allows handling significantly higher light levels. However, the consequences for a potential 3D imaging device need to be further investigated.\\
Furthermore, optically correlating light intensities via two-photon absorption in semiconductors is another promising technique to measure the inherent intensity correlations of thermal light \cite{boitier2009measuring}. While this method supports larger light levels, it has the same stability and alignment requirements of coherent interferometers.\\
While we have shown a proof of principle for 3D imaging with intensity interferometry, further investigations are required before a practical 3D imaging device can be built. Nevertheless, the advantages of intensity interferometry are quite compelling.

\section{Conclusion}
In this work, a novel 3D imaging technique that only exploits the bunching signature of real thermal light is implemented. Raster-scanning is used to measure the second-order correlation function of real thermal light reflected from a simple 3D object, on a point-by-point basis. This enables producing a topographical map of the object's 3D surface. Despite low photon count rates, reasonable measurement times of $250\,\text{s}$ per pixel are achieved. The measurement scheme utilizing gated SPAD technology to synchronize photon counting makes this possible. Further simulations show that the gated technique improves the measured SNR when including more light modes in the experiment. Using SPAD arrays instead of only a single detector pixel can further enhance the performance of the data acquisition as many more light modes could be correlated. With recent developments in SPAD technology and signal processing speed, even faster, larger sized, and cheaper SPAD detector systems might be available soon.\\
The presented technique leverages the temporal coherence properties of thermal sources to perform ToF-based 3D imaging. In principle, the method can be used with any thermal light source, including ambient sources such as the sun, so that it may be possible to make ToF measurements passively, without any hardware control of illumination sources. In addition, the presented gated photon-counting technique can be interesting for stellar intensity interferometers. In particular, when performing intensity interferometry with unbalanced photon count rates in the two interferometer arms, the gated technique can leverage the correlation histogram acquisition. This is conceivable when using telescopes with differently sized light collectors, as it will be the case for the Cherenkov Telescope Array \cite{pareschi2013status}.

\bibliography{sources}

\begin{thebibliography}{10}
\newcommand{\enquote}[1]{``#1''}

\bibitem{foix2011lock}
S.~Foix, G.~Alenya, and C.~Torras, \enquote{Lock-in time-of-flight (tof)
  cameras: a survey,} {\protect\JournalTitle{IEEE Sens. J.}} \textbf{11}
  (2011).

\bibitem{morimoto2020megapixel}
K.~Morimoto, A.~Ardelean, M.-L. Wu, A.~C. Ulku, I.~M. Antolovic, C.~Bruschini,
  and E.~Charbon, \enquote{Megapixel time-gated spad image sensor for 2d and 3d
  imaging applications,} {\protect\JournalTitle{Optica}} \textbf{7}, 346--354
  (2020).

\bibitem{Willo3DCam17}
F.~Willomitzer and G.~H\"{a}usler, \enquote{Single-shot 3d motion picture
  camera with a dense point cloud,} {\protect\JournalTitle{Opt. Express}}
  \textbf{25}, 23451--23464 (2017).

\bibitem{Harendt14}
B.~Harendt, M.~Gro{\ss}e, M.~Schaffer, and R.~Kowarschik, \enquote{3d shape
  measurement of static and moving objects with adaptive spatiotemporal
  correlation,} {\protect\JournalTitle{Appl. Opt.}} \textbf{53}, 7507--7515
  (2014).

\bibitem{antipa2018diffusercam}
N.~Antipa, G.~Kuo, R.~Heckel, B.~Mildenhall, E.~Bostan, R.~Ng, and L.~Waller,
  \enquote{Diffusercam: lensless single-exposure 3d imaging,}
  {\protect\JournalTitle{Optica}} \textbf{5}, 1--9 (2018).

\bibitem{Huber11}
F.~Huber, O.~Arold, F.~Willomitzer, S.~Ettl, and G.~H{\"a}usler, \enquote{{3D
  body scanning with "Flying Triangulation"},} in \emph{Proceedings of the
  $112^{th}$ DGaO Conference, P18 (2011).}, .

\bibitem{lidarSelfDriving}
J.~Levinson, J.~Askeland, J.~Becker, J.~Dolson, D.~Held, S.~Kammel, J.~Z.
  Kolter, D.~Langer, O.~Pink, V.~Pratt, M.~Sokolsky, G.~Stanek, D.~Stavens,
  A.~Teichman, M.~Werling, and S.~Thrun, \enquote{Towards fully autonomous
  driving: Systems and algorithms,} in \emph{2011 IEEE Intelligent Vehicles
  Symposium (IV),}  (2011), pp. 163--168.

\bibitem{GerdWhy}
G.~H{\"a}usler, C.~Faber, F.~Willomitzer, and P.~Dienstbier, \enquote{Why
  can’t we purchase a perfect single shot 3d-sensor?} in \emph{Proceedings of
  the $113^{th}$ DGaO Conference (2012).}, .

\bibitem{Caulier:10}
Y.~Caulier, \enquote{Inspection of complex surfaces by means of structured
  light patterns,} {\protect\JournalTitle{Opt. Express}} \textbf{18},
  6642--6660 (2010).

\bibitem{lange2001solid}
R.~Lange and P.~Seitz, \enquote{Solid-state time-of-flight range camera,}
  {\protect\JournalTitle{IEEE Journal of quantum electronics}} \textbf{37},
  390--397 (2001).

\bibitem{stoppa2010range}
D.~Stoppa, N.~Massari, L.~Pancheri, M.~Malfatti, M.~Perenzoni, and L.~Gonzo,
  \enquote{A range image sensor based on 10$\mu$m lock-in pixels in 0.18$\mu$m
  cmos imaging technology,} {\protect\JournalTitle{IEEE journal of solid-state
  circuits}} \textbf{46}, 248--258 (2010).

\bibitem{yasutomi2013time}
K.~Yasutomi, T.~Usui, S.~Han, M.~Kodama, T.~Takasawa, K.~Kagawa, and
  S.~Kawahito, \enquote{A time-of-flight image sensor with sub-mm resolution
  using draining only modulation pixels,} in \emph{Proc. 2013 Int. Image Sensor
  Workshop,}  (2013), pp. 357--360.

\bibitem{li2019mega}
F.~Li, F.~Willomitzer, P.~Rangarajan, and O.~Cossairt, \enquote{Mega-pixel
  time-of-flight imager with ghz modulation frequencies,} in
  \emph{Computational Optical Sensing and Imaging,}  (Optical Society of
  America, 2019), pp. CTh2A--2.

\bibitem{steel1983interferometry}
W.~H. Steel, \emph{Interferometry}, vol.~1 (CUP Archive, 1983).

\bibitem{liu2019non}
X.~Liu, I.~Guill{\'e}n, M.~La~Manna, J.~H. Nam, S.~A. Reza, T.~H. Le,
  A.~Jarabo, D.~Gutierrez, and A.~Velten, \enquote{Non-line-of-sight imaging
  using phasor-field virtual wave optics,} {\protect\JournalTitle{Nature}}
  \textbf{572}, 620--623 (2019).

\bibitem{willomitzer2019high}
F.~Willomitzer, F.~Li, M.~M. Balaji, P.~Rangarajan, and O.~Cossairt,
  \enquote{High resolution non-line-of-sight imaging with superheterodyne
  remote digital holography,} in \emph{Imaging and Applied Optics 2019 (COSI,
  IS, MATH, pcAOP),}  (Optical Society of America, 2019), p. CM2A.2.

\bibitem{willomitzer2019synthetic}
F.~Willomitzer, P.~V. Rangarajan, F.~Li, M.~M. Balaji, M.~P. Christensen, and
  O.~Cossairt, \enquote{Synthetic wavelength holography: An extension of
  gabor's holographic principle to imaging with scattered wavefronts,}
  {\protect\JournalTitle{arXiv preprint arXiv:1912.11438}}  (2019).

\bibitem{ulku2018512}
A.~C. Ulku, C.~Bruschini, I.~M. Antolovi{\'c}, Y.~Kuo, R.~Ankri, S.~Weiss,
  X.~Michalet, and E.~Charbon, \enquote{A 512$\times$ 512 spad image sensor
  with integrated gating for widefield flim,} {\protect\JournalTitle{IEEE
  Journal of Selected Topics in Quantum Electronics}} \textbf{25}, 1--12
  (2018).

\bibitem{ceccarelli2018fully}
F.~Ceccarelli, G.~Acconcia, A.~Gulinatti, M.~Ghioni, and I.~Rech,
  \enquote{Fully integrated active quenching circuit driving custom-technology
  spads with 6.2-ns dead time,} {\protect\JournalTitle{IEEE Photonics
  Technology Letters}} \textbf{31}, 102--105 (2018).

\bibitem{brown1956test}
R.~H. Brown and R.~Q. Twiss, \enquote{A test of a new type of stellar
  interferometer on sirius,} {\protect\JournalTitle{Nature}} \textbf{178},
  1046--1048 (1956).

\bibitem{brown1956correlation}
R.~H. Brown, R.~Q. Twiss \emph{et~al.}, \enquote{Correlation between photons in
  two coherent beams of light,} {\protect\JournalTitle{Nature}} \textbf{177},
  27--29 (1956).

\bibitem{glauber1963quantum}
R.~J. Glauber, \enquote{The quantum theory of optical coherence,}
  {\protect\JournalTitle{Physical Review}} \textbf{130}, 2529 (1963).

\bibitem{hanbury1974angular}
R.~Hanbury~Brown, J.~Davis, and L.~Allen, \enquote{The angular diameters of 32
  stars,} {\protect\JournalTitle{Monthly Notices of the Royal Astronomical
  Society}} \textbf{167}, 121--136 (1974).

\bibitem{tan2016optical}
P.~K. Tan, A.~H. Chan, and C.~Kurtsiefer, \enquote{Optical intensity
  interferometry through atmospheric turbulence,}
  {\protect\JournalTitle{Monthly Notices of the Royal Astronomical Society}}
  \textbf{457}, 4291--4295 (2016).

\bibitem{davis1999sydney}
J.~Davis, W.~Tango, A.~Booth, T.~t. Brummelaar, R.~Minard, and S.~Owens,
  \enquote{The sydney university stellar interferometer—{I}. the instrument,}
  {\protect\JournalTitle{Monthly Notices of the Royal Astronomical Society}}
  \textbf{303}, 773--782 (1999).

\bibitem{pilyavsky2017single}
G.~Pilyavsky, P.~Mauskopf, N.~Smith, E.~Schroeder, A.~Sinclair, G.~T. van
  Belle, N.~Hinkel, and P.~Scowen, \enquote{Single-photon intensity
  interferometry (spiify): utilizing available telescopes,}
  {\protect\JournalTitle{Monthly Notices of the Royal Astronomical Society}}
  \textbf{467}, 3048--3055 (2017).

\bibitem{horch2013intensity}
E.~Horch, G.~Van~Belle, R.~Genet, and B.~Holenstein, \enquote{Intensity
  interferometry for the 21st century,} {\protect\JournalTitle{Journal of
  Astronomical Instrumentation}} \textbf{2}, 1340009 (2013).

\bibitem{dravins2015long}
D.~Dravins, T.~Lagadec, and P.~D. Nu{\~n}ez, \enquote{Long-baseline optical
  intensity interferometry-laboratory demonstration of diffraction-limited
  imaging,} {\protect\JournalTitle{Astronomy \& Astrophysics}} \textbf{580},
  A99 (2015).

\bibitem{kieda2019astro2020}
D.~B. Kieda, G.~Anton, A.~Barbano, W.~Benbow, C.~Carlile, M.~Daniel,
  D.~Dravins, S.~Griffin, T.~Hassan, J.~Holder \emph{et~al.},
  \enquote{Astro2020 white paper state of the profession: Intensity
  interferometry,} {\protect\JournalTitle{arXiv preprint arXiv:1907.13181}}
  (2019).

\bibitem{kieda2019augmentation}
D.~Kieda, V.~Collaboration, S.~LeBohec \emph{et~al.}, \enquote{Augmentation of
  veritas telescopes for stellar intensity interferometry,}
  {\protect\JournalTitle{arXiv preprint arXiv:1908.03095}}  (2019).

\bibitem{le2006optical}
S.~Le~Bohec and J.~Holder, \enquote{Optical intensity interferometry with
  atmospheric cerenkov telescope arrays,} {\protect\JournalTitle{The
  Astrophysical Journal}} \textbf{649}, 399 (2006).

\bibitem{weiss2018stellar}
S.~A. Weiss, J.~D. Rupert, and E.~P. Horch, \enquote{Stellar photon correlation
  detection with the southern connecticut stellar interferometer,} in
  \emph{Optical and Infrared Interferometry and Imaging VI,}  vol. 10701
  (International Society for Optics and Photonics, 2018), p. 107010X.

\bibitem{guerin2017temporal}
W.~Guerin, A.~Dussaux, M.~Fouch{\'e}, G.~Labeyrie, J.-P. Rivet, D.~Vernet,
  F.~Vakili, and R.~Kaiser, \enquote{Temporal intensity interferometry: photon
  bunching in three bright stars,} {\protect\JournalTitle{Monthly Notices of
  the Royal Astronomical Society}} \textbf{472}, 4126--4132 (2017).

\bibitem{zmija2020led}
A.~Zmija, P.~Deiml, D.~Malyshev, A.~Zink, G.~Anton, T.~Michel, and S.~Funk,
  \enquote{Led as laboratory test source for astronomical intensity
  interferometry,} {\protect\JournalTitle{Optics Express}} \textbf{28},
  5248--5256 (2020).

\bibitem{liu2014lensless}
X.-F. Liu, X.-H. Chen, X.-R. Yao, W.-K. Yu, G.-J. Zhai, and L.-A. Wu,
  \enquote{Lensless ghost imaging with sunlight,} {\protect\JournalTitle{Optics
  letters}} \textbf{39}, 2314--2317 (2014).

\bibitem{defienne2020quantum}
H.~Defienne, J.~Zhao, E.~Charbon, and D.~Faccio, \enquote{Quantum illumination
  imaging with a single-photon avalanche diode camera,}
  {\protect\JournalTitle{arXiv preprint arXiv:2007.16037}}  (2020).

\bibitem{boiko2009quantum}
D.~Boiko, N.~Gunther, N.~Brauer, M.~Sergio, C.~Niclass, G.~Beretta, and
  E.~Charbon, \enquote{A quantum imager for intensity correlated photons,}
  {\protect\JournalTitle{New Journal of Physics}} \textbf{11}, 013001 (2009).

\bibitem{mccarthy2009long}
A.~McCarthy, R.~J. Collins, N.~J. Krichel, V.~Fern{\'a}ndez, A.~M. Wallace, and
  G.~S. Buller, \enquote{Long-range time-of-flight scanning sensor based on
  high-speed time-correlated single-photon counting,}
  {\protect\JournalTitle{Applied optics}} \textbf{48}, 6241--6251 (2009).

\bibitem{zhu2012thermal}
J.~Zhu, X.~Chen, P.~Huang, and G.~Zeng, \enquote{Thermal-light-based ranging
  using second-order coherence,} {\protect\JournalTitle{Applied optics}}
  \textbf{51}, 4885--4890 (2012).

\bibitem{mehringer2017optical}
T.~Mehringer, S.~Oppel, and J.~von Zanthier, \enquote{An optical multimode
  fiber as pseudothermal light source,} {\protect\JournalTitle{Applied Physics
  B}} \textbf{123}, 200 (2017).

\bibitem{schneider2018simulating}
R.~Schneider, C.~Biernoth, J.~H{\"o}lzl, A.~Pscherer, and J.~von Zanthier,
  \enquote{Simulating the photon stream of a real thermal light source,}
  {\protect\JournalTitle{Applied optics}} \textbf{57}, 7076--7080 (2018).

\bibitem{valencia2004distant}
A.~Valencia, G.~Scarcelli, and Y.~Shih, \enquote{Distant clock synchronization
  using entangled photon pairs,} {\protect\JournalTitle{Applied Physics
  Letters}} \textbf{85}, 2655--2657 (2004).

\bibitem{jun2013distance}
X.~Jun-Jun, F.~Chen, H.~Xiao-Chun, Z.~Jian-Kang, and Z.~Gui-Hua,
  \enquote{Distance ranging based on quantum entanglement,}
  {\protect\JournalTitle{Chinese Physics Letters}} \textbf{30}, 100301 (2013).

\bibitem{fox2006quantum}
M.~Fox, \emph{Quantum optics: an introduction}, vol.~15 (OUP Oxford, 2006).

\bibitem{loudon2000quantum}
R.~Loudon, \emph{The Quantum Theory of Light} (OUP Oxford, 2000).

\bibitem{mandel1995optical}
L.~Mandel and E.~Wolf, \emph{Optical coherence and quantum optics} (Cambridge
  university press, 1995).

\bibitem{MPDSPAD}
M.~P. Devices, \enquote{Spec sheet pdm series,}
  \url{www.micro-photon-devices.com}.

\bibitem{wagner2020intensity}
F.~Wagner, \enquote{Intensity interferometry-based 3d imaging,} M.{S}. thesis
  (FAU Erlangen-Nuremberg, 2020).

\bibitem{schneider2018correlation}
R.~Schneider, \enquote{Correlation experiments and data evaluation techniques
  with classical light sources in space and time,} Ph.D. thesis (FAU
  Erlangen-Nuremberg, 2018).

\bibitem{schottky1918spontane}
W.~Schottky, \enquote{{\"U}ber spontane stromschwankungen in verschiedenen
  elektrizit{\"a}tsleitern,} {\protect\JournalTitle{Annalen der physik}}
  \textbf{362}, 541--567 (1918).

\bibitem{boitier2009measuring}
F.~Boitier, A.~Godard, E.~Rosencher, and C.~Fabre, \enquote{Measuring photon
  bunching at ultrashort timescale by two-photon absorption in semiconductors,}
  {\protect\JournalTitle{Nature Physics}} \textbf{5}, 267--270 (2009).

\bibitem{pareschi2013status}
G.~Pareschi, T.~Armstrong, H.~Baba, J.~B{\"a}hr, A.~Bonardi, G.~Bonnoli,
  P.~Brun, R.~Canestrari, P.~Chadwick, M.~Chikawa \emph{et~al.},
  \enquote{Status of the technologies for the production of the cherenkov
  telescope array (cta) mirrors,} in \emph{Optics for EUV, X-Ray, and Gamma-Ray
  Astronomy VI,}  vol. 8861 (International Society for Optics and Photonics,
  2013), p. 886103.

\end{thebibliography}

\bibliographyfullrefs{sources}


\ifthenelse{\equal{\journalref}{aop}}{%
\section*{Author Biographies}
\begingroup
\setlength\intextsep{0pt}
\begin{minipage}[t][6.3cm][t]{1.0\textwidth} 
  \begin{wrapfigure}{L}{0.25\textwidth}
    \includegraphics[width=0.25\textwidth]{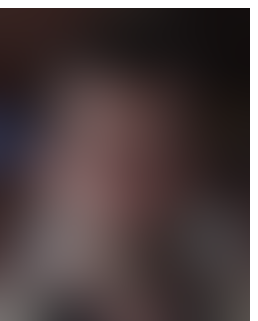}
  \end{wrapfigure}
  \noindent
  {\bfseries John Smith} received his BSc (Mathematics) in 2000 from The University of Maryland. His research interests include lasers and optics.
\end{minipage}
\begin{minipage}{1.0\textwidth}
  \begin{wrapfigure}{L}{0.25\textwidth}
    \includegraphics[width=0.25\textwidth]{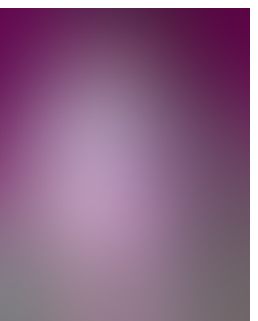}
  \end{wrapfigure}
  \noindent
  {\bfseries Alice Smith} also received her BSc (Mathematics) in 2000 from The University of Maryland. Her research interests also include lasers and optics.
\end{minipage}
\endgroup
}{}

\end{document}